\documentclass[aps,prd,preprint,showpacs]{revtex4}
\usepackage{amsmath,amssymb}
\usepackage{epsfig}
\def\ph{\phantom}
\def\beq{\begin{equation}}
\def\eeq{\end{equation}}
\def\hlf{\frac{1}{2}}

\def\G{\mathcal{G}}
\def\m{\tilde{m}}
\def\s{\sigma}
\def\I{\mathcal{I}}

\def\E{\mathcal{E}}
\def\A{\mathcal{A}}
\def\B{\mathcal{B}}
\def\C{\mathcal{C}}
\def\S{\mathcal{S}}
\begin{document}
\title{Strong field effects on binary systems in Einstein--aether theory}
\author{Brendan Z. Foster}
\email[]{B.Z.Foster@phys.uu.nl} \affiliation{Institute for
Theoretical Physics, Utrecht University, Leuvenlaan 4, NL-3584 CE
Utrecht, The Netherlands}
\date{September 23, 2008}
%
%-------------------------------------------------------
%
\begin{abstract}
``Einstein--aether" theory is a generally covariant theory of
gravity containing a dynamical preferred frame. This article
continues an examination of effects on the motion of binary pulsar
systems in this theory, by incorporating effects due to strong
fields in the vicinity of neutron star pulsars. These effects are
included through an effective approach, by treating the compact
bodies as point particles with nonstandard, velocity dependent
interactions parametrized by dimensionless ``sensitivities".
Effective post-Newtonian equations of motion for the bodies and
the radiation damping rate are determined. More work is needed to
calculate values of the sensitivities for a given fluid source;
therefore, precise constraints on the theory's coupling constants
cannot yet be stated. It is shown, however, that strong field
effects will be negligible given current observational
uncertainties if the dimensionless couplings are less than roughly
$0.01$ and two conditions that match the PPN parameters to those
of pure general relativity are imposed. In this case, weak field
results suffice. There then exists a one-parameter family of
Einstein--aether theories with ``small-enough" couplings that
passes all current observational tests.  No conclusion can be
reached for larger couplings until the sensitivities for a given
source can be calculated.
\end{abstract}
\pacs{04.50.+h, 04.30.Db, 04.25.Nx, 04.80.Cc}
\maketitle
%-----------------------------------------------------
%
%
\section{Introduction}
This article examines the motion of stellar systems in
``Einstein--aether" theory---an alternative theory of gravity that
permits breaking of Lorentz symmetry through a dynamical preferred
frame.  The general theory contains four dimensionless couplings
whose values can be constrained by comparing the predictions of
the theory with observations---in particular, observations of
binary pulsar systems. It will be demonstrated that all current
tests will be passed by  a one-parameter family whose couplings
are ``small-enough"---that is, on the order of $0.01$ or less.
Identifying whether there is a viable extension of this family to
large coupling values requires additional work beyond this
article.

The question of whether the physical world is exactly Lorentz
invariant has received increasing attention in recent years. This
interest is sourced largely by hints of Lorentz violation in
popular candidates for theories of quantum gravity---for instance,
string theory~\cite{Kostelecky:1988ta}, loop quantum
gravity~\cite{Gambini:1998it}, and noncommutative field
theory~\cite{Hewett:2000zp}.  More broadly, challenging the rule
of Lorentz symmetry means challenging the fundamentals of all of
modern physics, and doing that is just plain exciting.

The review~\cite{Mattingly:2005re} discusses a wide variety of
theoretical models that feature Lorentz-symmetry violating
effects, and observational searches for violations. So far no
conclusive sign of  Lorentz variance has been identified, and very
strong bounds exist on the size of couplings for Lorentz-violating
effects in standard model
extensions~\cite{Mattingly:2005re,Bluhm:2005uj}. The effects of
Lorentz violation in a gravitational context, however, are not
covered by these bounds.

Einstein--aether theory---or ``ae-theory" for short---is a
classical metric theory of gravity that contains an additional
dynamical vector field. The vector field ``aether" is constrained
to be timelike everywhere and of fixed norm. The aether can be
thought of as a remnant of unknown, Planck-scale,
Lorentz-violating physics. It defines a preferred frame, while its
status as a dynamical field preserves diffeomorphism invariance.
The fixed norm, which can always be scaled to unity, ensures that
the aether picks out just a spacetime direction and removes
instabilities in the unconstrained theory~\cite{Elliott:2005va}.

Much of past work on ae-theory has focussed on placing
observational bounds on the values of the  four free parameters
$c_n$ appearing in the ae-theory action, Eqn.~\eqref{ACT6}.
Constraints have been derived from the rate of primordial
nucleosynthesis~\cite{Carroll:2004ai}, the rate of \^{C}erenkov
radiation~\cite{Elliott:2005va}, the requirements of stability and
energy positivity of linearized wave modes~\cite{Foster:2005dk},
and the parameterized post-Newtonian (PN) form of the
theory~\cite{Eling:2003rd,Graesser:2005bg,Foster:2005dk}. A
summary of these constraints was presented
in~\cite{Foster:2005dk}, where it was shown that they are met by a
large two-parameter subset of the original four-parameter class of
theories.

Additional constraints on the $c_n$ come from  observations of
binary pulsar systems.  Study of the predictions of ae-theory for
binary pulsars was begun in~\cite{Foster:2006az}. There, an
expression for the rate of radiation damping in $N$-body systems
was derived to lowest non-trivial PN order and \textit{neglecting
effects due to strong fields in the vicinity of the bodies}.  It
was shown that a one-parameter subset of the two-parameter family
allowed by the collected constraints discussed
in~\cite{Foster:2005dk} would pass tests from binary pulsar
systems if there were justification for ignoring strong field
effects.  That neglect is dangerous, though, since the fields
inside neutron star pulsars should be very strong. Justification
requires an unclear assumption on the values of the $c_n$.

In this article, I will incorporate strong  field effects on
binary pulsar systems, calculating the PN equations of motion and
the rate of radiation damping of a system of strongly
self-gravitating bodies.  The effects will be handled via an
effective approach in which the compact bodies are treated as
point particles whose action contains nonstandard couplings that
depend on the velocity of the particles in the preferred frame.
The effective approach to $N$-body dynamics in relativistic
gravity theories has previously been employed in pure general
relativity (GR) and other alternative theories; see for example
~\cite{Einstein:1938yz,Eardley:1977a,WillEardley:1977a,Damour:1995kt,Goldberger:2004jt,Porto:2006bt}.
The new interactions are parametrized by dimensionless
coefficients, or ``sensitivities", whose values can be calculated
for a given stellar source by matching the effective theory onto
the exact, perfect fluid theory. Prior work ~\cite{Foster:2006az}
reveals just the form of the ``first" sensitivity at lowest order
in the self-potential of a body.

The expressions obtained can be used to constrain the allowed
class of ae-theories.  Observations of binary pulsar systems allow
for measurement of ``post-Keplerian" (PK) parameters that describe
perturbations of the binary's Keplerian orbit due to relativistic
effects.  These parameters are mostly ``quasi-static" ones, whose
expressions can be derived from the non-radiative parts of the PN
forms of the gravitational fields and the effective equations of
motion for the bodies. In addition, there is the radiation damping
rate, whose expression depends on the radiative parts of the
fields. The ae-theory expressions for the PK parameters differ
from those of pure GR in that they depend on the $c_n$, the
sensitivities, and the center-of-mass velocity of the system of
bodies.  Stating precise constraints for general $c_n$ values will
require work beyond the scope of this article---specifically, what
is needed is a method for dealing with dependence on the
unmeasurable center-of-mass velocity and a calculation of the
values of the sensitivities of a given source.

For the time being, a few comments can be made, which will be
defended below.  A crucial piece of information learned by
comparing the weak field limit of the effective theory with the
weak field limit of the perfect fluid theory~\cite{Foster:2006az}
is that the sensitivities will be ``small".  That is, they will be
at least as small as $(G_N m/d)^2$, where $m$ is the body's mass
and $d$ its size, times a $c_n$ dependent coefficient that must
scale at least linearly with $c_n$ in the small $c_n$ limit. For
neutron stars in pure GR, $(G_N m/d)\sim(0.1 \sim 0.3)$; it is
reasonable to expect something similar in ae-theory based on
studies of stellar solutions~\cite{Eling:2006df} and the fact that
ae-theory is generally ``close" in the small-$c_n$ limit, to GR
plus a non-dynamical vector field.

It then follows that bounds on the magnitude of violations of the
strong equivalence principle~\cite{Stairs:2003eg} constrain the
$c_n$ dependent factor to be less than $(0.01)(G_N m/d)^{-2}.$ It
further follows that the strong field corrections fall below the
level of current observational uncertainties when $|c_n| \lesssim
0.01$ and the two conditions that match the ae-theory PPN
parameters to those of GR are imposed.  Thus, weak field
analysis~\cite{Foster:2006az} suffices for small enough $c_n$, and
implies the existence of a one-parameter family of theories that
passes all current tests from binary pulsar systems.

I will now present the strong field formulas.  First, the
effective particle action is constructed, and the exact field
equations are defined in Sec.~\ref{sec2}.  The PN expansions of
the metric and aether fields are then given in Sec.~\ref{sec3},
and used to express the PN equations of motion for a binary system
in Sec.~\ref{sec4}. The rate of radiation damping is then
determined in Sec.~\ref{sec5}. Comments on dealing with
center-of-mass velocity and sensitivity dependence are given in
Sec.~\ref{sec6}, along with the argument for the viability of the
weakly coupled family of theories.

I follow the  conventions of Wald~\cite{Wald:1984rg}. In particular,
I use units in which the flat space speed-of-light $c = 1$, and I
use metric signature ${(-,+,+,+)}$.  This signature is opposite to
that employed in ~\cite{Foster:2006az}, but it is much more
convenient for calculations involving a time-space decomposition.
The ae-theory action is defined here in such a way as to permit easy
comparison between~\cite{Foster:2006az} and this article. The
following shorthand conventions  for combinations of the $c_n$ will
be used:
\begin{gather}
    c_{14} = c_1 + c_4,\\
    c_{123} = c_1 + c_2 + c_3,\\
    c_{\pm} = c_1 \pm c_3.
\end{gather}

When covariant equations are expanded  in Minkowskian coordinates,
the following conventions are observed.  Spatial indices will be
indicated by lowercase Latin letters from the middle of the
alphabet: $i,j,k,\dots$.  One exception is when the coefficients
$c_{1,2,3,4}$ are referred to collectively as $c_n$.   Indices
will be raised and lowered with the flat metric $\eta_{ab}$.
Repeated spatial indices will be summed over, regardless of
vertical position: $T_{ii} = \sum_{i = 1\dots 3} T_{ii}$.  Time
indices will be indicated by a $0$; time derivatives will be
denoted by an overdot: $\dot{f}\equiv\partial_0 f$.
%
%---------------------------------------------------
%---------------------------------------------------
%
\section{Effective action and field equations}\label{sec2}
\subsection{Particle action}
The aim of this work is to treat within ae-theory a system of
compact bodies that potentially possess strong internal
gravitational fields. The complicated internal workings of the
bodies will be dealt with via an effective approach that
reproduces the bulk motion of the bodies and the fields far from
them. Each body will be treated as a point particle with the
composition dependent effects encapsulated in nonstandard
couplings in the particle action.

The form of the effective action can be deduced from the following
considerations. The one-particle action $S_A$ will have the rough
form $S_A = -\tilde{m} \int dt \mathcal{O}$, where the integral is
along the particle worldline parametrized by $t$, $\tilde{m}$ has
dimensions of mass, and $\mathcal{O}$ is a sum of dimensionless
local scalar quantities. The fundamental theory has only one
dimensionful parameter $G$.  For a first approximation, the spin of
the body can be neglected. Derivative couplings in the particle
theory are then suppressed by powers of $(d/R)$, where $d$ is the
size of the underlying finite-sized body and $R$ is the radius of
curvature of the background spacetime. In addition, $S_A$ presumably
reduces to the standard free particle action if the particle is
comoving with the local aether and must be invariant under
reparametrization of the particle worldline.

These considerations imply the following one-particle action:
\beq\label{1bodyS}
    S_A=-\tilde{m}_A \int d\tau_A \big(1 + \s_A( u^a v_a+1) +
    \frac{\s'_A}{2}(u^a v_a+1)^2 + \cdots \big),
\eeq
where $A$ labels the body, $\tau_A$ is the proper time along the
body's curve, $v^a$ is the body's unit four-velocity, and $u^a$ is
the aether.  The quantity $u^a v_a$ expressed in a PN expansion
with the aether purely timelike at lowest order, is of order
$v^2$, the square of the velocity of the body in the aether frame.
By assumption, $v^2$ is first PN order (1PN). The 1PN corrections
to Newtonian equations of motion will follow from the part of the
action that is $m_A \times(\text{2PN})$, so  only the terms in
$S_A$ written above are needed for current purposes.  For a system
of $N$ particles, the action is given by the sum of $N$ copies of
$S_A$.

This action can be thought of as a Taylor expansion of the standard
worldline action, but with a mass that is a function of $\gamma
\equiv -u^a v_a$:
\beq
    S_A = -\int d\tau \,\tilde{m}_A[\gamma].
\eeq
The expansion is made about $\gamma=1$.  The parameters $\s,\s'$ are
then defined as
%
%check signs
\beq
    \s_A = -\frac{d \ln \tilde{m}_A}{d \ln \gamma}|_{\gamma=1},\quad
    \s'_A = \s_A+ \s_A^2 + {\bar\s}_A,\quad
    \bar{\s}_A = \frac{d^2 \ln \tilde{m}_A}{d(\ln\gamma)^2}|_{\gamma = 1}.
\eeq
This form of $S_A$ suggests that that $\s_A,\bar{\s}_A$ can be
determined by considering asymptotic properties of perturbations
of static stellar solutions.
%
%------------------------------
%
\subsection{Field equations}
The full action is the four-parameter ae-theory action $S$
\beq\label{ACT6}
    S = \frac{1}{16\pi G}\int d^4x\,\sqrt{|g|}\,\Big(
        R - K^{ab}_{\ph{ab}cd}\nabla_a u^c \nabla_b u^d
        +\lambda(u^a u^b g_{ab} + 1)\Big),
\eeq
plus the sum of $N$ copies of $S_A$~\eqref{1bodyS}, retaining only
the terms explicitly written above. Here,
\beq
        K^{ab}_{\phantom{ab}cd} = \big(c_1 g^{ab}g_{cd}
        +c_2\delta^a_c\delta^b_d + c_3\delta^a_d\delta^b_c
        - c_4 u^a u^b g_{cd}\big).
\eeq
While the sign of the $c_4$ term looks awkward, it permits easier
comparison with the results of~\cite{Foster:2006az}.

The field equations are then as follows. There are the Einstein
equations
\beq
    G_{ab} - S_{ab} = 8\pi G T_{ab},
\eeq
where
\beq
    G_{ab} = R_{ab} -\hlf R g_{ab},
\eeq
\beq
\begin{split}
    S_{ab} = &\nabla_c\bigl(K_{(a}^{\ph{(a}c}u_{b)} -
        K^c_{\ph{c}(a} u_{b)} - K_{(ab)}u^c\bigr)\\&
        +c_1\bigl(
            \nabla_a u_c\nabla_b u^c
            - \nabla_c u_a\nabla^c u_b\bigr)
       +c_4(u^c\nabla_c u_a)( u^d\nabla_d u_b)\\
       &+\lambda u_a u_b +\frac{1}{2}g_{ab}
          (K^c_{\ph{a}d}\nabla_c u^d),
\end{split}
\eeq
 with
\beq
    K^a_{\phantom{a}c} = K^{ab}_{\phantom{ab}cd}\nabla_b u^d,
\eeq
and $T^{ab}$ is the particle stress tensor
\beq\label{Tfull}
    T^{ab} = \sum_A \tilde{m}_A\tilde{\delta}_A\,
    \bigl[A^1_{A} v_A^a v_A^b
    + 2A^2_{A} u^{(a} v_A^{b)} \bigr],
\eeq
with a covariant delta-function
\beq
    \tilde{\delta}_A = \frac{\delta^3(\vec{x} -
    \vec{x}_A)}{v_A^0 \sqrt{|g|}},
\eeq
and
\begin{gather}
    A^1_{A} =1+\s_A-\frac{\s'_A}{2}\big((u_c v_A^c)^2-1\big),\\
    A^2_{A} =-\s_A-\s'_A(u_c v_A^c + 1).
\end{gather}
The aether field equation is
\beq\label{AEQ6}
    \nabla_b K^{ba}
        = c_4 (u^c\nabla_c u_b)\nabla^a u^b + \lambda u^a
        + 8\pi G \sigma^a,
\eeq
where
\beq\label{sigmafull}
    \sigma^a =\sum_A \tilde{m}_A\tilde{\delta}_A\, A^2_{A} v_A^a.
\eeq
Varying $\lambda$ gives the constraint $g_{ab}u^a u^b = -1$.
Eqn.~\eqref{AEQ6} can be used to eliminate $\lambda$, giving
\beq
    \lambda = -u^a\Big(\nabla_b K^b_{\phantom{b}a}
    - c_4(\nabla_a u^b)(u^c\nabla_c u_b) - 8\pi G \sigma_a\Big).
\eeq
The covariant equation of motion for a single particle has the form
\beq\label{eqmofull}
    \nabla_b T_A^{ab} - \nabla_b\big((\sigma_A)^a u^b\big) - (\sigma_A)_b\nabla^au^b
    =0,
\eeq
where $T_A^{ab}$ and $(\sigma_A)^a$ are the one-particle summands
in~\eqref{Tfull} and~\eqref{sigmafull}. This can be written more
explicitly as
\beq
    v_A^b\nabla_b (A^1_{A} v_A^a + A^2_{A} u^a) - A^2_{A} v_{Ab}\nabla^a u^b = 0.
\eeq
%
%--------------------------------------------------
%
\section{Post-Newtonian expansion}
\subsection{Fields}\label{sec3}

The PN expansion of the fields can be determined by iteratively
solving the field equations in a weak field, slow motion
approximation~\cite{Will:2001mx,Foster:2005dk}.  A background of a
flat metric and constant aether is assumed, and a Lorentzian
coordinate system with the time direction defined by the
background aether is chosen. Following the procedures
of~\cite{Foster:2005dk} gives
\beq
\begin{split}
    g_{00} = & -1 + 2\sum_A \frac{G_N \tilde{m}_A}{r_A}
                - 2\sum_{A,B} \frac{G_N^2\tilde{m}_A\tilde{m}_B}
                        {r_Ar_B}
                 -2\sum_{A,B\neq A}\frac{G_N^2\tilde{m}_A\tilde{m}_B}
                        {r_A r_{AB}}\\
               &  + 3 \sum_{A}\frac{G_N \tilde{m}_A}{r_A} v_A^2
                            (1+\s_A),\\
    g_{ij} = & \Big(1
                + 2\sum_A \frac{G_N\tilde{m}_A}{r_A}\Big)\delta_{ij},\\
    g_{0i} =    & \sum_A B^-_A
                    \frac{G_N\tilde{m}_A}{r_A} v_A^i
               + \sum_A B^+_A
                    \frac{G_N \tilde{m}_A}{r_A^3}(v_A^j
                    r_A^j)r_A^i,
\end{split}
\eeq
where $r^i_A = x^i - x_A^i$, $r^i_{AB} = x^i_A - x^i_B$,
\beq
    B^{\pm}_A = \pm \frac{3}{2} \pm\frac{1}{4}(\alpha_1 -
        2\alpha_2)\Big(1 + \frac{(2-c_{14})}{(2c_+-c_{14})}\s_A\Big) -\frac{1}{4}(8+\alpha_1)\Big(1+\frac{c_-}{2c_1}\s_A\Big),
\eeq
and
\begin{gather}
    G_N = \frac{2}{2-c_{14}} \,G,\\
    \alpha_1 = -\frac{8(c_3^2 + c_1 c_4)}{2c_1 - c_+c_-},\label{alpha1}\\
    \alpha_2 = \frac{\alpha_1}{2} - \frac{(c_1+2c_3 - c_4)(2c_1 +
            3c_2+c_3+c_4)}{(2-c_{14})c_{123}}\label{alpha2}.
\end{gather}
The numerical values of the PPN parameters $\alpha_1$ and $\alpha_2$
are constrained to be very small by weak field experiments, via
analysis that allows for a possible lack of Lorentz symmetry in the
underlying theory~\cite{Will:2001mx}. There are two independent
pairs of conditions on the $c_n$ that will set $\alpha_1$ and
$\alpha_2$ to zero.  One pair is
\beq\label{CONDITION6}
    c_2 = -\frac{2c_1^2 + c_1c_3 - c^2_3}{3c_1},\qquad
    c_4 = -\frac{c_3^2}{c_1}.
\eeq
The other is $c_+ = c_{14} = 0$.  With this second pair, the
spin-1 and spin-0 wave speeds diverge (Sec.~\ref{sec5}); also, the
spin-0 linearized energy density vanishes while that of spin-1
remains finite~\cite{Eling:2005zq}. Observational signatures of
this behavior have not been worked out, and I will not consider
these conditions further here. Hence, the first pair of conditions
is assumed below whenever attention is restricted to the case of
vanishing $\alpha_{1}$ and $\alpha_2$.

The aether to order of interest is
\beq
\begin{split}
    u^0 = & 1 + \sum_A \frac{G_N \tilde{m}_A}{r_A},\\
    u^i = & \sum_A C^-_A \frac{G_N \tilde{m}_A}{r_A} (v_A)^i
            + \sum_A  C^+_A
                \frac{G_N \tilde{m}_A}{r_A^3}(v_A^j r_A^j) r_A^i,
\end{split}
\eeq
where
\beq
    C^{\pm}_A = \big(\frac{8+\alpha_1}{8c_1}\big)\big(c_- - (1-c_-)\s_A\big) \pm
        \frac{(2-c_{14})}{2}\Big(\frac{(\alpha_2 - \frac{\alpha_1}{2})}
                {(c_1 + 2c_3 - c_4)} + \frac{1}{c_{123}}\s_A\Big).
\eeq
%
%When $\alpha_1 = \alpha_2 = 0$, we have
%
%\begin{gather}
%    B_{1A} = -2(1+\frac{c_-}{2c_1}\s_A)\\
%    B_{2A} = -\frac{3}{2}\\
%    B_{3A} = \frac{1}{c_1}\big(c_- + (1+c_-)\s_A\big)\\
%    B_{4A} = \frac{3(2c_1 + c_+ c_-)}{2 c_+^2}\s_A
%\end{gather}
%
The results of this section are equivalent to the weak field expressions
obtained in~\cite{Foster:2005dk} when $\s_A$ is set to zero.
%
%---------------------------------------------------------
%--------------------------------------------------------
%
\subsection{Post-Newtonian equations of motion}\label{sec4}
The equations of motion for the system of compact bodies follow by
expressing the exact result~\eqref{eqmofull} in a PN expansion
using the forms of the fields given above.  The Newtonian order
result can be used to define the effective two-body coupling $\G$
and the ``active" gravitational mass $m$:
\beq
    \dot{v}_A^i = \sum_{B\neq A}  \frac{-G_N
    \tilde{m}_B}{(1+\sigma_A)r^3_{AB}}r^i_{AB}
         \equiv \sum_{B\neq A}  \frac{-\G_{AB} m_B}{r^3_{AB}}r^i_{AB},
\eeq
with the two-body coupling
\beq
    \G_{AB} = \frac{G_N}{(1+\sigma_A)(1+\sigma_B)},
\eeq
and the active gravitational mass
\beq\label{MASS}
    m_B = (1+\sigma_B)\tilde{m}_B.
\eeq
These definitions arise by requiring that $\G_{AB} =\G_{BA}$ and
that $m_B/ \tilde{m}_B$ depend on just $\sigma_B$.

Using the Newtonian result and  continuing with the expansion
leads to the 1PN equations of motion, expressed here just for the
case of a binary system:

\beq
\begin{split}
   \dot{v}^i_1 =& \frac{\G m_2}{r^2}{\hat{r}^i}
        \Big[-1
        + 4\frac{\tilde{m}_2}{r}
        + \Big(1-\frac{2}{1+\s_2}D\Big)\frac{\m_1}{r}\\
        &-\hlf\Big(2+3\s_1 +\frac{\s'_1}{1+\s_1}\Big) v_1^2
        -\Big(\frac{3}{2}(1+\s_2)+(E-D)\Big)v_2^2\\
        &- 2D v^j_1 v^j_2
        +3(E-D) (v^j_2\hat{r}^j)^2\Big]\\
        &+\frac{\G m_2}{r^2}\Big[v^i_1
        \Big(v^j_1\hat{r}^j\big(4+3\s_1-\frac{\s'_1}{1+\s_1}\big)
        -3(1+\s_1)v^j_2\hat{r}^j\Big)\\
        &+v^i_2(2D v^j_1\hat{r}^j
            - 2E v^j_2\hat{r}^j)\Big],
\end{split}
\eeq
where $\G = \G_{12}$, $r^i = r^i_1 - r^i_2$, and
\begin{gather}
    D = -\frac{1}{4}(8+\alpha_1)\Big(1+\frac{c_-}{2c_1}(\s_1 + \s_2)   +\frac{(1-c_-)}{2c_1}\s_1\s_2\Big),\\
     E = -\frac{3}{2} - \frac{1}{4}(\alpha_1 - 2\alpha_2)
        \Big(1 + \frac{(2-c_{14})}{(c_1 + 2c_3-c_4)}(\s_1 +\s_2)
            +\frac{(2-c_{14})}{2c_{123}}\s_1\s_2\Big)
\end{gather}
The expression for $\dot{v}^i_2$ is obtained by exchanging all
body-1 quantities and body-2 quantities, including the switch
$r^i \rightarrow - r^i$.

The ``Einstein--Infeld--Hoffman"
Lagrangian~\cite{Einstein:1938yz}---that is,  the effective
Lagrangian expressed purely in terms of particle quantities---can
be determined by working backwards from the equations of motion.
It is
\beq
\begin{split}
    L =& -(m_1 + m_2) + \hlf(m_1 v_1^2 + m_2 v_2^2)\\ &+ \frac{1}{8}
        \bigg(\Big(1-\frac{\s'_1}{1+\s_1}\Big)v_1^4 +
        \Big(1-\frac{\s'_2}{1+\s_2}\Big)v_2^4\bigg)\\
        &+\frac{\G m_1 m_2}{r}\bigg[1 + \frac{3}{2}\Big(
            (1+\s_1)v_1^2 + (1+\s_2)v_2^2\Big)\\
            &-\hlf\Big(\frac{\G m_1}{r}(1+\s_2)+\frac{\G
            m_2}{r}(1+\s_1)\Big)\\
            &+ D(v_1^j v_2^j) + E (v_1^j \hat{r}^j v_2^k
            \hat{r}^k)\bigg].
\end{split}
\eeq
This Lagrangian is not Lorentz invariant unless $\s_A = \s'_A =
0$.  This follows from the analysis of Will~\cite{Will:1993ns} and
the list of criteria therein.  In particular, the action and the
equations of motion depend on the velocity of the system's center of
mass in the aether frame.
%
%%---------------------------------------------------------------
%---------------------------------------------------------
%
\section{Radiation damping rate}\label{sec5}
The radiation damping rate is the rate at which the particle system
loses energy via gravity-aether radiation. This energy loss
manifests as a change in the orbital period of a binary system,
equating the energy radiated to minus the change in mechanical
energy. The expression for the rate in the effective particle theory
can be determined by adapting the methods of~\cite{Foster:2006az},
which were used to find the rate for a system of weakly
self-gravitating perfect fluid bodies in ae-theory. It will be
convenient to introduce the parameter $s_A$
\beq
    s_A = \sigma_A/(1+\sigma_A),
\eeq
and to work
with the active gravitational mass~\eqref{MASS}
\beq
    m_A = (1+\s_A)\tilde{m}_A = \tilde{m}_A/(1-s_A).
\eeq
\subsection{Wave forms}
The method of~\cite{Foster:2006az} begins by assuming a background
of a flat metric and constant aether, with a coordinate system
with respect to which the background metric is the Minkowski
metric $\eta_{ab}$ and the background aether is aligned with the
time direction. The metric and aether perturbations are then
decomposed into irreducible transverse and longitudinal pieces.
The spatial vectors $u^i$ and $h_{0i}$ are written as:
\beq
    h_{0i} = \gamma_i + \gamma_{,i}\quad
    u^i = \nu^i + \nu_{,i},
\eeq
with $\gamma_{i,i} = \nu^i_{,i} = 0$.  The spatial metric $h_{ij}$
is decomposed into a transverse, trace-free tensor, a transverse
vector, and two scalar quantities giving the transverse and
longitudinal traces:
\beq
    h_{ij} = \phi_{ij} + \hlf P_{ij}[f] + 2\phi_{(i,j)} + \phi_{,ij},
\eeq
where
\beq
    0=\phi_{ij,j} = \phi_{jj} = \phi_{i,i},
\eeq
and
\beq
    P_{ij}[f] = \delta_{ij}f_{,kk} - f_{,ij};
\eeq
hence, $P_{ij}[f]_{,j} = 0$, and $h_{ii} = ( f + \phi)_{,ii}$.
Further, define
\beq
    F = f_{,jj}.
\eeq
The list of variables then consists of a transverse-traceless
spin-2 tensor $\phi_{ij}$, transverse spin-1 vectors
$\gamma_i,\nu^i,\phi_i$, and spin-0 scalars $\gamma,\nu,F,\phi,
h_{00}$, and $u^0$.  The Lorentz gauge,  or any obvious extension
of it, does not usefully simplify the ae-theory field equations.
Instead, the following convenient conditions will be imposed:
\beq
    0 = u^i_{,i} = h_{0i,i} = h_{i[j,k]i},
\eeq
or equivalently,
\beq\label{GGE6}
    0 = \nu = \gamma = \phi_i.
\eeq
Because $\phi_i$ is transverse, these constitute just four
conditions.

Following~\cite{Foster:2006az}, the field equations can then be
linearized and expressed in terms of the above variables, and
sorted to obtain a set of wave equations with matter terms and
nonlinear terms as sources. Having done this, the linear
contributions can be seen by inspection to satisfy a conservation
law.  This fact implies the existence of a conserved source
$\tau^{ab}$
\beq
        \tau^{ab} = T^{ab} - \sigma^a \delta^b_0 +
        \tilde{\tau}^{ab},
\eeq
where $T^{ab}$ and $\sigma^a$ are as defined in
Eqns.~\eqref{Tfull} and~\eqref{sigmafull}, and $\tilde{\tau}^{ab}$
is constructed from nonlinear terms---its precise form will not be
needed.   The non-symmetric $\tau^{ab}$ satisfies the conservation
law with respect to the right-index only:
$\tau^{ab}_{\phantom{ab},b} = 0$.  The corresponding conserved
total energy $E$ and momentum $P^i$ to lowest PN order are
\begin{gather}\label{TOTAL}
    E = \int d^3 x\,\tau^{00} = \sum_A \tilde{m}_A = \sum_A (1-s_A)
    m_A,\\
    P^i = \int d^3 x\,\tau^{i0} =\sum_A m_A v_A^i.
\end{gather}
Conservation of $P^i$ means that the system center-of-mass $X^i$
defined via $m_A$
\beq
    X^i = \frac{\sum_A m_A x_A^i}{\sum_A m_A},
\eeq
is unaccelerated to lowest order.

The field equations reduce to the following.  For spin-2,
\beq
    \frac{1}{w_2^2}\ddot{\phi}_{ij} - \phi_{ij,kk} = 16\pi G
    \tau_{ij}^{\rm TT},
\eeq
where $\rm TT$ signifies the transverse, trace-free components, and
\beq
    w_2^2 = \frac{1}{1-c_+}.
\eeq
For spin-1,
\begin{gather}
    \frac{1}{w_1^2}\big(\ddot{\nu^i} + \ddot{\gamma_i}\big)
        = \frac{16\pi G}{2c_1 - c_+ c_-}
            (c_+ \tau_{i0} + (1-c_+)\sigma^i)^{\rm T},\\
    (c_+ \nu^i + \gamma_i)_{,kk} = -16 \pi G \tau_{i0}^{\rm T},
\end{gather}
where $\rm T$ signifies the transverse components, and
\beq
    w_1^2 = \frac{2c_1 - c_+ c_-}{2(1-c_+)c_{14}}.
\eeq
For the spin-0 variables, the constraint gives to linear order
\beq
    u^0 = 1+\frac{1}{2}h_{00}.
\eeq
Non-linear corrections to this are of uninteresting order, as explained in more
detail in~\cite{Foster:2006az}.  The other equations are
\begin{gather}
    \frac{1}{w_0^2} \ddot{F} - F_{,kk} =
        \frac{16 \pi G c_{14}}{2-c_{14}} \big( \tau_{kk}
        - \frac{2+3c_2+c_+}{c_{123}}\tau^{\rm L}_{kk}
        +\frac{2}{c_{14}}\tau_{00}\big),\\
    (F - c_{14} h_{00})_{,kk} = -16\pi G \tau_{00},\\
    (1+c_2)\dot{F}_{,i} + c_{123} \dot{\phi}_{,kki}
        = -16\pi G \tau^{\rm L}_{i0},
\end{gather}
where $\rm L$ signifies the longitudinal component, and
\beq
    w_0^2 = \frac{(2-c_{14})c_{123}}{(2+3c_2+c_+)(1-c_+)c_{14}}.
\eeq

All these equations can be solved formally via Greens function
methods, and the resulting integrals expanded in a far field, slow
motion approximation.  The expressions can be further simplified
using the conservation of $\tau^{ab}$.  A result that holds within
the approximation scheme is that for a field $\psi$ satisfying a
wave equation with speed $w$ evaluated at field point $x^i \equiv
|x| \hat{n}^i$ with only outgoing waves,
\beq
    w \psi_{,i}  = - \dot{\psi} \hat{n}^i.
\eeq
Also, differentially transverse becomes equivalent to geometrically
transverse to $\hat{n}^i$.

The results to lowest PN order and ignoring static contributions are
as follows. For spin-2,
\beq
    \phi_{ij} = \frac{2G}{|x|} \ddot{Q}_{ij}^{\rm TT},
\eeq
where the right-hand side is evaluated at time $(t - |x|/w_2)$ and
the quadrupole moment $Q_{ij}$ is the trace-free part of the
system's second mass moment $I_{ij}$:
\beq
    I_{ij} = \sum_A m_A x_A^i x_A^j.
\eeq
For spin-1 variables,
\begin{gather}\label{one}
    \nu^i = \frac{-2G}{|x|}\frac{1}{2c_1 - c_+
    c_-}\Big(\frac{\hat{n}^j}{w_1}(\frac{c_+}{1-c_+}
        \ddot{Q}_{ij}+\ddot{\mathcal Q}_{ij})
            - 2 \Sigma^i\Big)^{\rm T},\\
    \gamma_i = -c_+ \nu^i,
\end{gather}
where the right-hand side of the first equation is evaluated at
time $(t-|x|/w_1)$, $\mathcal{Q}_{ij}$ is the trace-free part of
the rescaled mass moment $\mathcal{I}_{ij}$:
\beq
    \mathcal{I}_{ij} = \sum_A s_A m_A x^i_A x^j_A,
\eeq
and
\beq\label{DefSig}
    \Sigma^i = -\sum_A s_A m_A v_A^i.
\eeq
For spin-0 variables,
\begin{gather}
\begin{split}
    F = \frac{-2G}{|x|}\frac{c_{14}}{2-c_{14}}
        \Big[&\Big(\big(\frac{2\alpha_2 - \alpha_1}{2(2c_+ -c_{14})}+3\big) \ddot{Q}_{ij}
            + \frac{2}{w_0^2 c_{14}} \ddot{\mathcal Q}_{ij}\Big)\hat{n}^i \hat{n}^j\\&
            +\frac{2\alpha_2 - \alpha_1}{2(2c_+ -c_{14})}\ddot{I}
            +\frac{2}{3w_0^2 c_{14}} \ddot{\mathcal I}
            - \frac{4}{w_0 c_{14}} \hat{n}^i \Sigma^i\Big],
\end{split}\\
    h_{00} = \frac{1}{c_{14}} F,\\
    \dot{\phi}_{,i} =- \frac{1+c_2}{c_{123}}\dot{f}_{,i},
\end{gather}
where the right-hand side of the first equation is evaluated at time
$(t-|x|/w_0)$,
and $I = I_{ii}$, $\mathcal{I} = \mathcal{I}_{ii}$.

At this point, the expected smallness of the sensitivities,
mentioned in the introduction, can be explained. One should take
the  weak field  limit ($s_A \rightarrow$ ``small") of the above
wave forms and compare them with the perfect-fluid theory wave
forms determined in~\cite{Foster:2006az}. The only
$s_A$-dependence that remains at potentially leading order is in
$\Sigma_i$.  Comparing~\eqref{DefSig} with Eqn.~(85)
of~\cite{Foster:2006az} indicates that in the small $s_A$ limit,
\beq\label{weaks}
    s_A = (\alpha_1 - \frac{2}{3}\alpha_2)
            \frac{\Omega_A}{m_A} + \mathcal{O}(\frac{G_N m}{d})^2,
\eeq
where $\Omega_A$ is the binding energy of the body: $\Omega/m \sim
(G_N m/d)$, where $d$ is the characteristic size of the body. The
implication is that when $\alpha_1 = \alpha_2 = 0$, $s$ must scale
as $(G_N m/d)^2$, times a $c_n$ dependent coefficient. This
coefficient should scale at least linearly in $c_n$, in the
$c_n\rightarrow 0$ limit, to ensure finiteness of the
perturbations.
\subsection{Damping rate expression}
For the next step, the wave forms are inserted into an expression for the rate of
change of energy $\dot{\mathcal{E}}$.  This expression can be derived via the Noether
charge method of Iyer and Wald~\cite{Wald:1993nt,Iyer:1994ys}, using
the ae-theory Noether charges derived in~\cite{Foster:2005fr}, with the result:
\beq
    \dot{\mathcal{E}} = \frac{-1}{16\pi G}\int d\Omega R^2
        \Big(\frac{1}{2w_2} \dot{\phi}_{ij}\dot{\phi}_{ij}
        +\frac{(2c_1 - c_+c_-)(1-c_+)}{w_1}\dot{\nu}^i\dot{\nu}^i
        +\frac{2-c_{14}}{4 w_0 c_{14}} \dot{F}\dot{F}\Big)
        + \dot{O}, \eeq
where $\dot{O}$ is a total time-derivative that will be argued
away in a moment.

Using the above results for the wave forms, performing the angular
integral, and ignoring $\dot{O}$ gives
\beq\label{NRG6}
    \dot{\mathcal{E}} = - G_N(
        \frac{\A_1}{5} \dddot{Q}_{ij}\dddot{Q}_{ij}
        + \frac{\A_2}{5} \dddot{Q}_{ij}\dddot{\mathcal Q}_{ij}
        + \frac{\A_3}{5} \dddot{\mathcal Q}_{ij}\dddot{\mathcal Q}_{ij}
        + \B_1 \dddot I \dddot I
        + \B_2 \dddot I \dddot{\mathcal I}
        + \B_3 \dddot{\mathcal I} \dddot{\mathcal I}
    + \C \dot{\Sigma}^i\dot{\Sigma}^i),
\eeq
where
\beq\label{COEFFS}
    \A_1 = \big(1-\frac{c_{14}}{2}\big)\bigg(\frac{1}{w_2}
        + \frac{2 c_{14}c^2_+}{(2c_1-c_+c_-)^2}\frac{1}{w_1}
        +\frac{c_{14}}{6(2-c_{14})}\Big(3
            +\frac{2\alpha_2 - \alpha_1}{2(2c_+ -c_{14})}\Big)^2
                \frac{1}{w_0}\bigg),
\eeq
\beq
    \A_2 = \bigg(\frac{(2-c_{14})c_+}{2c_1-c_+c_-}\frac{1}{w_1^3}
        + \Big(1+\frac{2\alpha_2 - \alpha_1}
                {6(2c_+ -c_{14})}\Big)\frac{1}{w_0^3}\bigg),
\eeq
\beq
    \A_3 = \frac{1}{c_{14}}\Big(\frac{2-c_{14}}{4}\frac{1}{w_1^5} -
            \frac{1}{3}\frac{1}{w_0^5}\Big),
\eeq
\beq
    \B_1 =\frac{c_{14}}{72}\Big(\frac{2\alpha_2 - \alpha_1}
        {2(2c_+ -c_{14})}\Big)^2\frac{1}{w_0},
\eeq
\beq
    \B_2 = \frac{2\alpha_2 - \alpha_1}{12(2c_+ -c_{14})}
            \frac{1}{w_0^3},
\eeq
\beq
    \B_3 = \frac{1}{6c_{14}}\frac{1}{w_0^5},
\eeq
\beq
    \C = \frac{2}{3 c_{14}}\Big(\frac{2-c_{14}}{w_1^3} +
    \frac{1}{w_0^3}\Big).
\eeq
The coefficients $\A_1$, $\B_1$, and $\C$ are respectively
identical to  $\mathcal{A}$, $\B$, and $\C$
of~\cite{Foster:2006az}.   Taking the weak field limit corresponds
to retaining only the $\A_1$, $\B_1$, and $\C$ terms, and invoking
the relation~\eqref{weaks} for $s_A$ in the $\C$ term. In the case
that $\alpha_1 = \alpha_2 = 0$, $\B_1$ vanishes, as does $s_A$ in
the weak field limit. The weak field damping rate in this case
then contains only a quadrupole contribution and is identical to
the GR rate when $\A_1 = 1$. This remaining curve of $c_n$ values
intersects the range of values allowed by collected constraints
considered in~\cite{Foster:2005dk}, as illustrated in Figure~1.
Thus, this curve gives a one-parameter family of viable
ae-theories if the weak field results alone are sufficient.
%
%
%------------------------------------------------------------
\begin{figure}[b]
\centering \epsfig{file=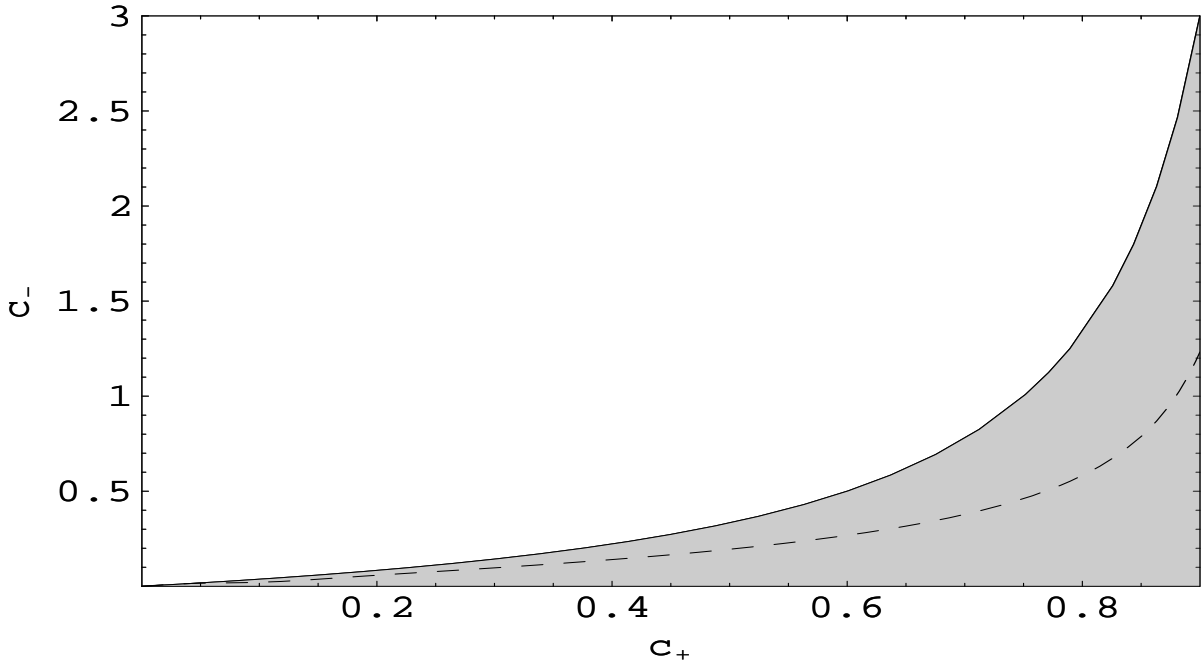, width=\textwidth,
height=3.5in} \caption{Class of allowed ae-theories, if strong
field effects in binary pulsar systems can be ignored. The
four-dimensional $c_n$ space has been restricted to the
$(c_+,c_-)$ plane by setting the PPN parameters $\alpha_1$ and
$\alpha_2$ to zero via the conditions~\eqref{CONDITION6}.  The
shaded region is the region allowed by primordial nucleosynthesis,
\^{C}erenkov radiation, linearized stability and energy
positivity, and PPN constraints, demarcated
in~\cite{Foster:2005dk}. The dashed curve is the curve along which
binary pulsar tests will be satisfied, assuming ae-theory weak
field expressions. Specifically, it is the curve along which
$\mathcal{A}_1 =1$ in the $\alpha_{1}=\alpha_2 = 0$ case, so that
the damping rate~\eqref{NRG6} is identical to the quadrupole
formula of general relativity. Along both this curve and the
boundary of the allowed region, $c_-\rightarrow \infty$ as $c_+
\rightarrow 1$. The curve remains within the allowed region for
all $c_+$ between 0 and 1. As explained in Sec.~\ref{sec6}, strong
field effects may lead to system dependent corrections to the
binary pulsar curve for large $c_n$; however, all such curves will
coincide with the weak field curve for $|c_n| \lesssim 0.01$ given
current observational uncertainties.}
\end{figure}
%
%
%-------------------------------------------

To simplify the expression~\eqref{NRG6}, it is crucial to note that
the damping rate is calculated to lowest PN order using the
Newtonian results for the motion of the system. Thus, the system's
motion can be decomposed into a uniform center-of-mass
motion---recall the conservation of $P^i$---and a fixed Keplerian
orbit in the center-of-mass frame. Since the motion is steady-state,
the damping rate must have no secular time dependence.  This
observation implies that secular terms in $\dot\E$ arising from
$\dddot{\mathcal I}_{ij}$, see below, must cancel with secular terms
in $\dot O$.  In addition, the non-secular portion of $\dot O$ will
average to zero when a time average of the damping rate over an
orbital period is taken, since it is a total time derivative. Thus,
$\dot O$ can be discarded.

Hence restricting attention to a binary system, and taking a time
average over an orbital period, the expression reduces as follows.
First, define the quantities
\beq
    m = m_1 + m_2,\quad \mu_A = m_A/m,\quad \mu = m_1 m_2/m,
\eeq
and the vectors
\begin{gather}
    r^i = x_1^i - x_2^i,\quad v^i = \dot{r}^i,\\
     X^i = \mu_1 x^i_1 + \mu_2 x^i_2,\quad V^i = \dot{X}^i.
\end{gather}
To Newtonian order, $\dot{v^i} = -(\G m/r^2) \hat{r}^i$, and
$\dot{V}^i = 0$.  $I_{ij}$ can be diagonlized
\beq
    I_{ij} = \mu r^i r^j + m X^i X^j,
\eeq
so that
\beq
    \dddot{I}_{ij} = \frac{2\G\mu m}{r^2}(3 \hat{r}^i \hat{r}^j\dot{r}
    - 4 v^{(i}\hat{r}^{j)}).
\eeq
As for $\I_{ij}$,
\beq
    \I_{ij} = \mu (s_1 \mu_2 + s_2 \mu_1) r^i r^j + m (s_1 \mu_1 +
    s_2\mu_2)X^i X^j + 2\mu(s_1 - s_2)r^{(i}X^{j)},
\eeq
and
\beq
    \dddot{\I}_{ij} = S \dddot{I}_{ij} - 6V^{(i} \dot{\Sigma}^{j)}
            + 2\mu(s_1-s_2)\dddot{r}^{(i}X^{j)},
\eeq
where
\beq
    \S = s_1\mu_2 + s_2\mu_1,
\eeq
and
\beq
    \dot{\Sigma}_i = (s_1 - s_2)\frac{\G\mu m}{r^3}r^i.
\eeq
Terms in $\dddot{\I}_{ij}$ with $X^i$ dependence are
secular; following the discussion above, they can be discarded.

Substituting into Eqn.~\eqref{NRG6} and imposing the time average
gives the final expression
\begin{multline}\label{RADRATE}
    \dot{\mathcal E} = -G_N\Big< \Big(\frac{\G\mu m}{r^2}\Big)^2\\
        \times\Big[\frac{8}{15}
        (\A_1 + \S \A_2 + \S^2 \A_3)(12 v^2 - 11 \dot{r}^2)\\
        + 4(\B_1 + \S\B_2 + \S^2 \B_3)\dot{r}^2\\
        +(s_1 - s_2)^2\Big(\C + \frac{6}{5} (3 \A_3 V^2 + (\A_3 + 30
        \B_3)(V^i\hat{r}^i)^2)\Big)\\
        +(s_1-s_2)\Big(\frac{8}{5}(\A_2 + 2 \S \A_3)(3 v^i V^i - 2
        V^i\hat{r}^i v^j \hat{r}^j) + 12(\B_2 + 2\S\B_3)V^i\hat{r}^i v^j
        \hat{r}^j\Big)
        \Big]\Big>,
\end{multline}
where the angular brackets denote the time average.
%
%------------------------------------------------------
%------------------------------------------------------
%
\section{Observational constraints}
\label{sec6}
\subsection{Center-of-mass velocity dependence}
While the aether frame center-of-mass velocity $V^i$ of a binary
system is not directly measurable, dependence of a binary
systems's motion on $V^i$ should actually be beneficial for
constraining the theory.  This is because constraints arise from a
failure to observe $V^i$ dependent effects. It may be possible to
formulate such constraints without having to determine the
physical frame, as in the manner of bounds on the PPN parameter
$\alpha_{2}$. The presence of alignment between the sun's spin
axis and the ecliptic plane signals the absence of frame dependent
effects, and leads to a strong bound of $|\alpha_2| < 4\times
10^{-7}$~\cite{Will:2001mx}. This argument does require the
assumption that the component of the preferred frame in the sun's
rest frame is not conveniently aligned with the sun's spin axis;
such an assumption may generally be required for similar
arguments. For example, $V^i$ dependence should cause a binary's
orbital plane to precess, but not if $V^i$ happens to be normal to
the plane.

An assumption on the order of magnitude of the norm $V$ is necessary
to justify the use of just the leading PN order expressions for the
PK parameters when applied to observed binary systems.  The validity
of the 1PN expressions depends on whether corrections of relative
order $v^2$ and $(V^4/v^2)$ are smaller than observational
uncertainties. Terms of order $v^2$ are negligible for all observed
systems, for now, although the ``double pulsar"~\cite{Kramer:2006nb}
is pushing this limit.  For all but the double pulsar, $v^2\sim
10^{-6}$, and uncertainties are at least a thousand times
this~\cite{Stairs:2003eg}. The double pulsar PSR J0737-3039A/B is
the so-far unique binary containing two pulsars. The orbital
velocity is high, $v^2 \sim 10^{-5}$, and the presence of two
pulsars happens to make measurement of system parameters much easier
and thus more precise---the smallest relative uncertainty is
$10^{-4}$ on the rate of periastron advance. The $v^2$ corrections
are therefore small enough for now, but it is expected that
precision will increase to probe the next PN order within the next
10-20 years~\cite{Stairs:2003eg}.

The $V^i$ dependent terms must feature $c_n$ dependent factors,
since it is known that there is no center-of-mass velocity
dependence at next PN order in pure GR~\cite{Will:2001mx}. Ignoring
those factors for the moment, validity of leading PN order for the
double pulsar requires that $(V^4/v^2) \lesssim 10^{-4}$, giving
$V^2 \lesssim 10^{-4.5}$, or $(V^2/v^2) \lesssim 10^{0.5}\approx 3$.
For other systems, given uncertainties ranging from
$(10^{-1}\sim10^{-3})$, the conditions are $(V^4/v^2) \lesssim
(10^{-1}\sim10^{-3})$, giving $V^2 \lesssim (10^{-3.5} \sim
10^{-2.5})$, or $(V^2/v^2) \lesssim (10^{2.5}\sim 10^{1.5})\approx
(300 \sim 30)$. Presumably, the $c_n$ dependent factor actually goes
to zero as some positive power of $c_n$, so $V$ can be larger in the
small $c_n$ limit.  A reasonable first guess for the aether frame is
the rest frame of the cosmic microwave background. A typical
velocity for compact objects in our galaxy in this frame is $V^2\sim
10^{-6}$, so the restriction on $V$ is met.
\subsection{Constraints in the small coupling regime}
A formula for the sensitivities for a given source should be
obtainable by comparing the strong field results of this article
with analogous results in the exact perfect fluid theory. Higher
order terms in the exact theory must be calculated, though, since
the leading order results of~\cite{Foster:2006az} only give the
$O(G_N m/d)$ part of $s$ expressed in~\eqref{weaks}. The
calculation can be done in the case of a single body that is
static except for a constant aether frame velocity, by, for
example, continuing the iterative procedure used to determine the
PPN parameters~\cite{Will:1993ns,Foster:2005dk}. The process may
be lengthy, but straightforward.

I have shown that the sensitivity of a body will scale with the
body's self-potential like $\beta[c_n](G_N m/d)^2$, where $\beta$
is some $c_n$-dependent coefficient that scales at least as fast
as $c_n$ in the small $c_n$ limit.  Even in the absence of a
formula for the sensitivities and precise knowledge of
center-of-mass velocities, two useful comments can be derived.
First, a constraint can be roughly stated: $|\beta|
\lesssim(0.1\sim 1)$. Second, there exists a one-parameter family
of theories that passes all current constraints, obtained by
restricting to $c_n$ with magnitude less than roughly $0.1$ and
imposing the two PPN conditions, the one weak field damping rate
condition, and the collected non-binary conditions.

The condition that $|\beta| \lesssim (0.1\sim 1)$ follows from
constraints~\cite{Stairs:2003eg} on the magnitude of violations of
the strong equivalence principle---that is, that a body's
acceleration is independent of its composition.  A violation would
lead to a polarization of the orbit of pulsar systems due to
unequal acceleration of the binary bodies in the gravitational
field of the galaxy.  The observed lack of polarization in neutron
star--white dwarf systems leads to a constraint that can be stated
here as $s < 0.01$, where here $s$ is the sensitivity of the
neutron star in the considered pulsars. Assuming that $(G_N m/d)
\approx (0.1\sim 0.3)$ for the pulsar, as it is in GR, the
constraint on the size of $\beta$ arises. It is possible that when
the weak field conditions are imposed, $\beta$ will automatically
satisfy the above inequality; certainly it will in the small $c_n$
regime when $|c_n| < 0.01$.

The statement, that current tests  will be satisfied if the weak
field conditions are imposed and the remaining degree of $c_n$
freedom satisfies $|c_n| \lesssim 0.01$, can be derived by
considering the battery of binary pulsar tests. First, consider
tests that probe only the quasi-static PK parameters---that is,
all but the damping rate. The tightest quasi-static test comes
from the double pulsar~\cite{Kramer:2006nb}. The relative size of
the strong field corrections to the weak field expressions will be
$O(s_A)$, while the prediction of GR has been confirmed to within
a relative observational uncertainty of $0.05\%$.  Requiring $s
\lesssim 10^{-3}$ and assuming that $(G_N m/d) \approx (.1\sim
.3)$ for the pulsars, the condition $|c_n| \lesssim 0.01$ arises.
Given this and the two conditions that set the PPN parameters
$\alpha_1$ and $\alpha_2$ to zero, all current quasi-static tests
will be passed.

Tests that incorporate the damping rate will  also be satisfied by
the small-$c_n$ condition and the weak field conditions.  I note
first that for systems in which the damping rate is probed,
uncertainty on its measurement dominates uncertainties on
quasi-static parameters~\cite{Will:2001mx,Stairs:2003eg}. Thus, it
is conventional  to use the measurements of the quasi-static
parameters to solve for the mass values of the binary bodies. When
$\alpha_{1} =\alpha_2 = 0$, and $|c_n| \lesssim 0.01$, so that the
expressions for the quasi-static parameters are close to those of
GR, the predicted mass values will also be close.

Now, the dipole contribution to $\dot\E$ can be significant in
asymmetric systems where the sensitivity of one body is much larger
than the other.  The dipole contribution is
\beq
    \dot{\E}_{Dipole} = -G_N\big< (\frac{\G\mu m}{r^2})^2\big>
            \C (s_1 - s_2)^2,
\eeq
which is of order $(\C s^2/ 10 v^2)$ compared to the quadrupole
and monopole contribution, where $s$ is the dominant sensitivity.
An applicable system is a neutron star--white dwarf binary, since
for a typical white dwarf, $(G_N m/d) \sim 10^{-3}$.  Constraints
have been derived~\cite{Stairs:2003eg} on the magnitude of dipole
radiation from neutron star--white dwarf binaries PSR B0655+64 and
PSR J1012+5307  by requiring that the dipole radiation rate be no
larger than the observed rate. The analysis applied here leads to
the condition $\C s^2 \lesssim 10^{-4}$, where $s$ is the
sensitivity of the neutron star. In the small $c_n$ regime, this
translates again to the condition $|c_n| \lesssim 0.01$.

For double neutron star binaries,  the dipole rate is further
suppressed by the similarities of the sensitivities, and the
quadrupole and monopole contributions become dominant. The
tightest test involving radiation is associated with the
Hulse--Taylor binary PSR1913+16, with a relative uncertainty of
$0.2\%$~\cite{Will:2001mx,Stairs:2003eg}.  In the small $c_n$
regime, the condition $\A_1 = 1$  matches the leading order
damping rate to that of GR. The strong field corrections are of
relative order $s$; to be smaller than the uncertainty again
requires $|c_n| \lesssim 0.01$.

This upper limit on $|c_n|$ will decrease as observational
uncertainties decrease. The most promising candidate for lowering
the limit is the double pulsar: 2PN-order and spin-dependent effects
should be observable within the next ten or twenty
years~\cite{Kramer:2006nb}. Another type of system, yet undetected,
for which high levels of accuracy could be obtained is a neutron
star--black hole binary, as the structureless black hole would
decrease noise due to finite-size effects and mass transfer between
the bodies.

For $|c_n| > 0.01$, strong field contributions to the expressions
for the PK parameters may be significant.  Those contributions for
a given source cannot yet be calculated, so the theory cannot be
checked against observations.  Thus, there is no conclusion yet on
the viability of large $c_n$ values.  If it were possible to
calculate precise predictions for a given binary system, then each
observed system would imply an extension from small to large $c_n$
of the curve of allowed values. The only physically viable values
would be those for which the curves for all observed systems
overlapped within error.
%
%---------------------------------------------------
%------------------------------------------------
%
%
\begin{acknowledgments}
I wish to thank Alessandra Buonanno, Cole Miller, Ira Rothstein,
Clifford Will, and especially Ted Jacobson for fruitful
discussions. This research was supported in part by the NSF under
grant PHY-0601800 at the University of Maryland.
\end{acknowledgments}
%
%--------------------------------------------------------
%-----------------------------------------------------
%
%\bibliography{MyBib}
%

%
%------------------------------------------------------------------
%-------------------------------------------------------------------
%
\end{document}